\documentclass[apj]{emulateapj}
\usepackage{subfigure}
\usepackage{amsmath}

\begin{document}

\title{On the IMF in a triggered star formation context$^\dagger$}
\author{Tingtao Zhou\altaffilmark{1,2}, Chelsea X. Huang\altaffilmark{1,3},
 D.N.C. Lin\altaffilmark{1,4,5}, Matthias Gritschneder\altaffilmark{4,6},
 \& Herbert Lau\altaffilmark{7} }
\affil{$^1$Kavli Institute for Astronomy \& Astrophysics and
School of Physics, Peking University, Beijing China, edmondztt@gmail.com}
\affil{$^2$Department of Physics, MIT, USA}
\affil{$^3$Department of Astrophysical Sciences, Princeton University, USA}
\affil{$^4$UCO/Lick Observatory, University of California, USA}
\affil{$^5$Institute for Advanced Studies, Tsinghua University, Beijing, China.}
\affil{$^6$University Observatory Munich, Germany}
\affil{$^7$Argelander Institute, University of Bonn, Germany}

\keywords{stars: formation-ISM: individual objects: Pipe Nebula -
HII regions-ISM: clouds-ISM: structure - methods: analytical
}

\begin{abstract}
The origin of the stellar initial  mass function (IMF) is a fundamental
issue in the theory of star formation. It is generally fitted with a
composite power-law. Some clues on the progenitors can be found in
dense starless cores which have a core mass function
(CMF) with similar shape. In the low-mass end, these mass functions increase
with mass, albeit the sample may be somewhat incomplete, whereas they decrease with
mass in the high-mass end.  There is an offset in the turn-over
mass between the two mass distributions. The stellar mass for the IMF peak is
lower than the corresponding core mass for the CMF peak in the Pipe
nebula by about a factor of three. Smaller offsets are found between the IMF and the CMFs
in other nebulae. We suggest that the offset is likely to be induced during a
star-burst episode of global star formation which is triggered by the formation
of a few O/B stars in the multi-phase media, which is naturally emerged by the
onset of a thermal instability in the cloud-core formation process.
We consider the scenario that the ignition of a few massive stars photoionizes the
warm medium between the cores, increases the external pressure, reduces their
Bonnor-Ebert mass and triggers the collapse of some previously stable
cores. We quantitatively reproduce the IMF in the
low-mass end with the assumption of additional rotational fragmentation.
\end{abstract}

\section{Introduction}
Recent infrared measurements of dust extinction, as well as CO and
${\rm NH_3}$ maps of the filaments in molecular clouds reveal a population of embedded
cores mostly confined by the global external pressure from the
inter-core gas. These cores are closely associated with the
progenitors of young stellar objects. In such filaments there are
gravitationally bound cores (or pre-stellar cores),
and embedded protostars \citep{Andre2010}. But, in the Pipe
nebula nearly all the cores are starless \citep{Lada2008}.
Only the most massive cores in Pipe are gravitationally bound and
might collapse. The shape of the CMF of these dense
cores \citep{Lada2008} appears to be qualitatively similar to the broken power-law
of the stellar IMF \citep{Kroupa2002,Kroupa2011}.
An in-depth analysis of their structure and evolution may be useful
for the construction of a star formation theory.

One noticeable difference between the CMF and the IMF is an offset between the two distributions.
The ratio of the characteristic mass of the IMF over the characteristic mass of the CMF is
generally smaller than unity. This ratio ranges from $\sim1:3$ in the Pipe nebula
\citep{Lada2009} to $>1:2$ in the Orion \citep{Nutter2007}
and Aquila nebulae \citep{Konyves2010}. \citet{Lada2009} proposed that
this offset is due to a direct one-to-one mapping from the cores to the
stars with sufficient amount of mass loss during the star formation process.
It is hard to explain in this scenario how the low-mass stars are produced by the
originally stable low-mass cores.

Many studies modeling the origin of IMF highlight factors such as the
accretion rate of protostars, turbulent fragmentation and accretion of cores
\citep[see e.g.,][]{Bonnell2006,Clark2008,Anathpindika2011}.
In the meantime, constraints such as the small age
spread of stars in young clusters are inconsistent in these models.

We propose a scenario to
explain the transition from the CMF to the IMF as well as a synchronized star formation,
triggered by a few first formed O/B
stars in the nebula.
Many authors \citep[e.g.,][]{Stromgren,Spitzer1978} studied the effects
of UV radiation from massive stars onto the surrounding regions.
The UV radiation generates a hot ionized region (Str{\"o}mgren sphere),
increasing the ambient pressure. Consequently, an isothermal shock is
driven through the nebula. Influenced by the Str{\"o}mgren sphere, denser
sub-structures are rapidly enhanced and most of the pre-existing
pressure confined cores are compressed \citep[e.g.,][]{Gritschneder2009MNRAS}.
The ionization timescale is generally short compared to the
hydrodynamical timescales, such that the increase in the background
pressure and temperature can be considered instant in most cases.

In this work, we develop a quantitative understanding of the
consequences of this sudden change in the ambient pressure and
temperature in the context of the transition from the CMF to the IMF.
A direct result is the reduction of the critical Bonnor-Ebert mass \citep{Ebert1955,Bonnor1956}, making
previous unbound cores gravitationally unstable. Therefore, the whole
region experiences a rapid star burst, synchronizing star formation.
The shifted Bonnor-Ebert mass also naturally leads to the transition
from the CMF distribution to the IMF distribution.

We base our calculations on a Pipe-like cloud, with a starless CMF
as the starting point of the evolution. We investigated the formation of
a starless CMF in our previous work (\citealt{Huang2013}, hereafter H13).
In H13, we suggest that, prior to the onset of global star formation,
the cores and the inter-core gas are two separate phases in pressure
equilibrium, which possibly result from thermal instability \citep{Lin2000}.
The dynamics of coagulation \citep{Murray1996} and ablation
\citep{Murray1993} dominate the evolution of the system and the star
formation timescale is prolonged by the turbulent pressure or magnetic
pressure inside the cloud \citep{LazarianVishniac99}.

In this paper, we consider the consequences of the ignition of
the first massive stars in a starless nebula.  In \S2, we suggest
that the photoionization of the inter-cloud gas leads to a
decrease in the critical Bonnor-Ebert mass and triggers
a global burst of star formation.  We use this reduced critical
mass and the starless CMF from the Pipe to obtain the transition to a
stellar IMF in \S3. We assume different star formation statistics,
showing that uncertainties such as binary mass ratio distribution do
not affect the IMF shape in a certain parameter range. The discussion
and conclusions are presented in \S4.

\section{Triggered star formation}
After the formation of the first massive stars, their stellar
luminosity has a significant influence on the subsequent evolution
of the surrounding media. In the presence of a strong UV flux from
nearby O/B stars, the diffuse neutral medium, originally at 100~K,
hereafter referred to as the warm medium, would be ionized and
heated up, while the dense molecular region, originally at 10~K,
hereafter referred to as the cold medium, would still be mostly
shielded. The resulting increase in the external pressure leads to a
reduction in the Bonnor-Ebert mass of the cold cores and a decrease
in the star formation timescale.

\subsection{The ionization of the warm medium\label{sec:ionization of warm media}}
The ionizing photons from the native stars create Str{\"o}mgren
spheres in the surrounding warm medium. For a
typical O/B star with an effective UV photon ($E_\gamma>13.6\,{\rm eV}$)
emission rate of $Q_0\sim 10^{48\sim49}\,{\rm s^{-1}}$, the Str{\"o}mgren
radius $R_S$ is determined by
\begin{equation}
Q_0 \sim \frac{4\pi}{3}\, R_S^3 \,\alpha_B\, n_{\rm warm}^2,
\end{equation}
where $\alpha_B = 2.6\times10^{-13}\,T_4^{0.833-0.034\log{T_4}}\,{\rm cm^3s^{-1}}$ is
the recombination coefficient \citep{Bruce2011}. We define $T_4$ as
the temperature of the warm medium in units of $10^4$\,K, $Q_{\rm 0,49}$ as
the number of UV photons in units of $10^{49}\,{\rm s}^{-1}$, and
$n_{\rm warm,2}$ as the warm medium hydrogen number density in units
of $10^2\,{\rm cm^{-3}}$. Within
$R_S\sim 3\,{\rm pc}\,Q_{\rm 0,49}^{\frac{1}{3}}\,n_{\rm warm,
  2}^{-\frac{2}{3}}$ of the ionizing sources, the warm medium
is nearly fully ionized, with $T_{\rm warm}\sim 10^4\,{\rm K}$. This
new temperature corresponds to an increase by a factor of 10$\sim$100.

More realistic calculations of equilibrium temperature have been
performed with version 08.00 of CLOUDY (last described by
\citealt{Ferland1998}). The effective temperature of the central star is
taken to be 34,700~K with a surface luminosity of log$(L/L_\odot)=5.59$.
This is a typical value for a  main sequence star with a mass of
$40M_{\sun}$, produced with the  Cambridge stellar evolution code
STARS \citep{Eggleton1971}. Assuming a blackbody, the star emits
$31\%$ of its power in hydrogen ionization photons, corresponding
to $Q_{0,49}\sim1$.

We assume a solar metallicity for the gas and a standard dust
composition. We use a constant hydrogen density of
$n_{\rm H}=774\,{\rm cm^{-3}}$ to simulate the warm medium in Pipe
nebula, to be consistent with the values in \citet{GritschnederLin2012}.

The results show that there is a sharp temperature dropoff from
several thousand degrees to several hundred degrees at a particular
depth of the cloud. The solid line in Figure \ref{fig:CLOUDY(a)}
shows the temperature of the warm medium versus the distance from the
star. This indicates the Str{\"o}mgren radius is $0.7 \,{\rm pc}$
and the temperature inside the ionized sphere is around $9000 \,{\rm K}$.
Given the typical size of a nebula, which is usually a few pc, only a few massive
stars are needed to ionize most of the inter-core media, leading to a much higher
exterior pressure.

\subsection{Heating of the cold cores}
The ionizing UV flux can also penetrate into dense cores.
We first consider the heating from the radiation which balances the
cooling from recombination.

We denote $R_{\rm ion}$ to be the distance from an O/B star at which a
typical cold core is heated by the external radiation to 100~K. By
neglecting the loss of ionizing photons from the star through the
warm medium, we estimate $R_{\rm ion}$ from:


\begin{equation}
\frac{R_{\rm ion}}{0.7\,{\rm pc}} \sim
(\frac{Q}{10^{49}\,{\rm s^{-1}}})^{\frac{1}{2}}
\,(\frac{r_c}{0.1 \,{\rm pc}})^{-\frac{1}{2}}
\,(\frac{n_c}{10^4 \,\rm {cm^{-3}}})^{-1} \\
\,(\frac{T_c}{100 \,{\rm K}})^{-0.5},
\end{equation}

where
\begin{equation}
Q_0\,\frac{\pi \,r_c^2} {4\pi \,R_{\rm ion}^2}
\sim \frac{4\pi}{3}\, r_c^3 \,\alpha_{\rm B} \,n_{\rm cold}^2,
\end{equation}
with $\alpha_B$ here is the same as in section~\ref{sec:ionization of warm media}.

\begin{figure*}
\begin{center}
\subfigure[]
{
\epsscale{.40}
\centering
\includegraphics[width=0.46\linewidth]{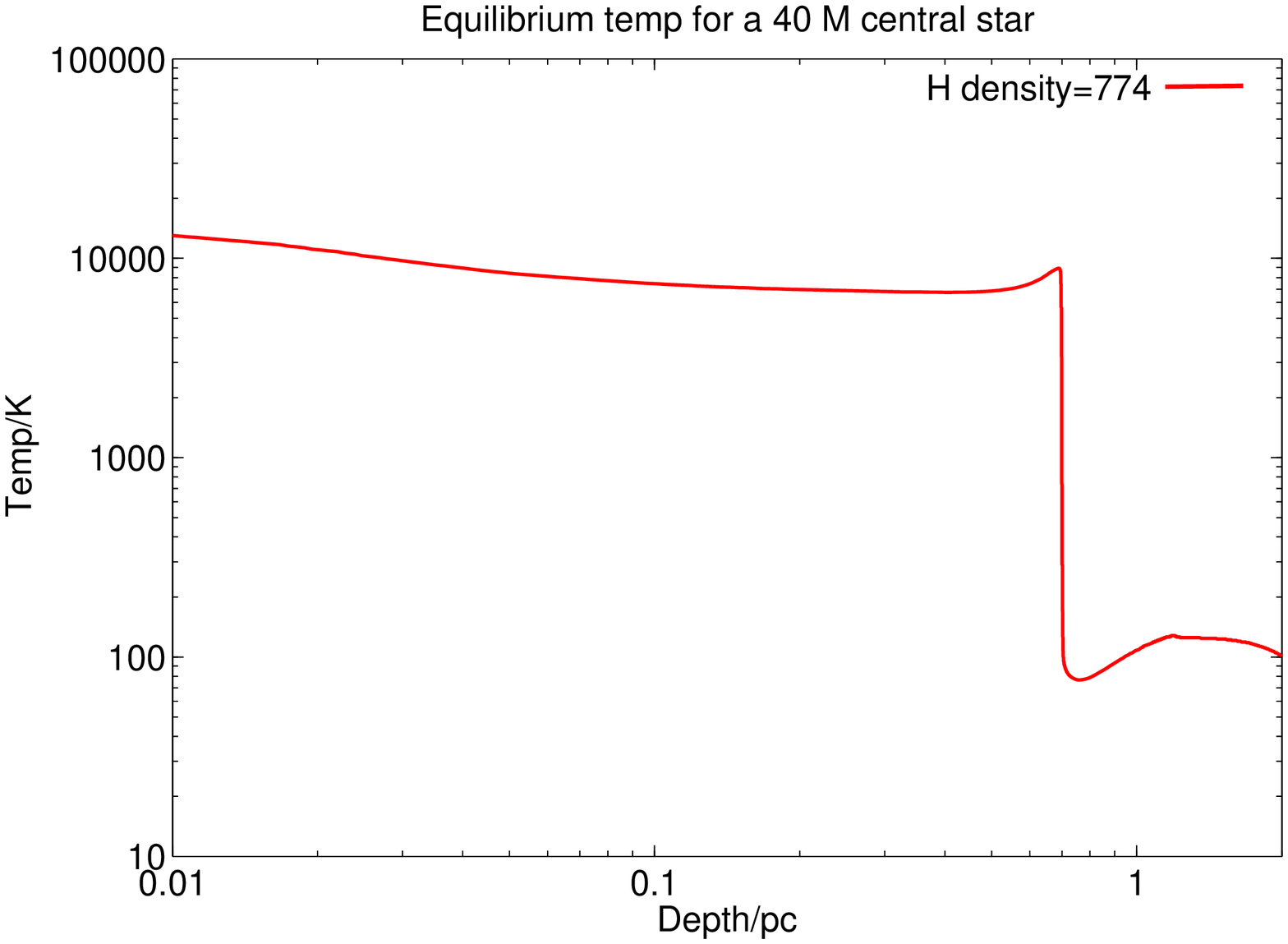}
\label{fig:CLOUDY(a)}
}
\subfigure[]
{
\epsscale{.40}
\centering
\includegraphics[width=0.45\linewidth]{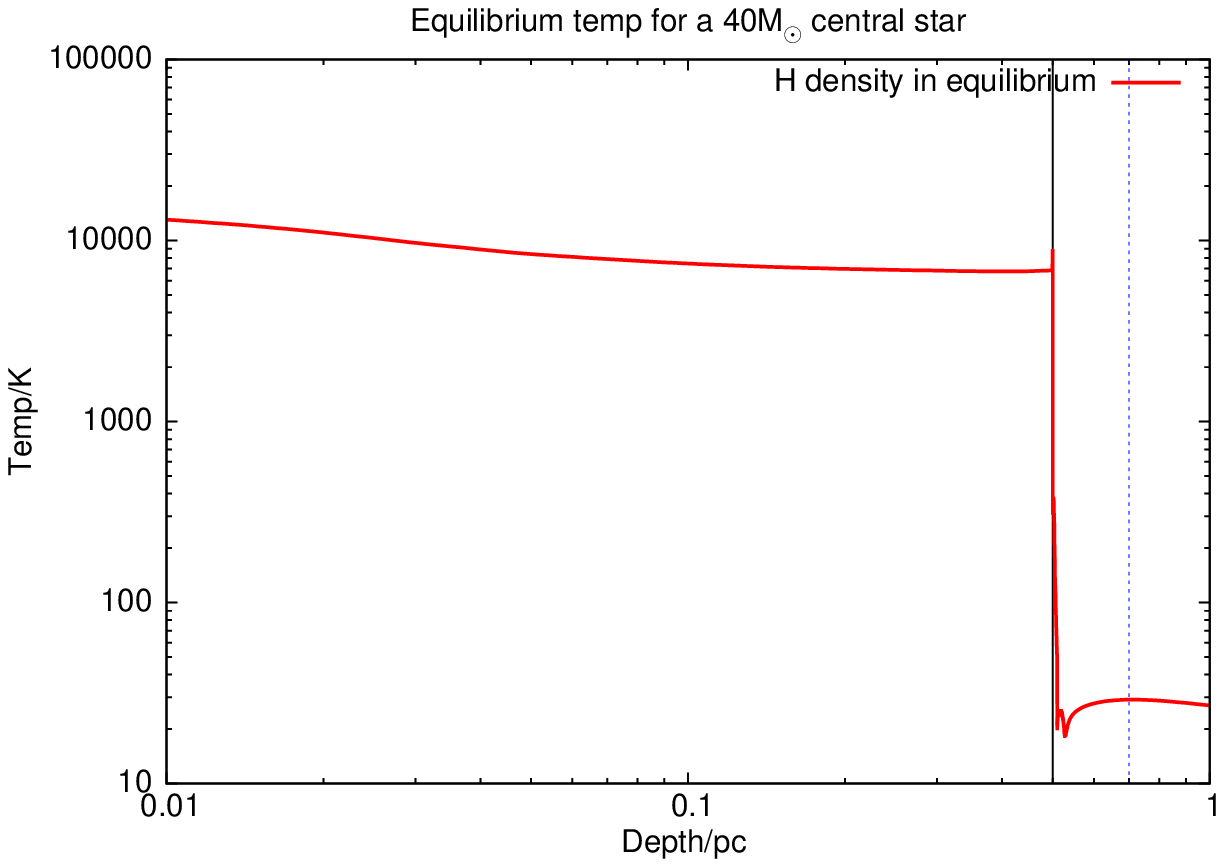}
\label{fig:CLOUDY(b)}
}
\end{center}
\caption{
Resulting temperature profiles in the surrounding of a 40 $M_{\sun}$ star for
different gas profiles.
(a) The red line displays the temperature profile in a medium with constant density
$n_0 =774\,{\rm cm^{-3}}$. The transition from ionized to un-ionized
medium  happens at $0.7\,{\rm pc}$ (the Str{\"o}mgen radius).
(b) This setup represents a cold core at a distance of $0.6\,{\rm pc}$ from the
central star. The initial
density profile is set to $n_0 =774\,{\rm cm^{-3}}$ (inter-core medium) inside $0.5\,{\rm pc}$
and $n_0 = 7300\,{\rm cm^{-3}}$ (core medium) beyond $0.5\,{\rm pc}$. The solid
vertical line indicates the inner side boundary of the core, i.e. the location of
the density jump. The dashed vertical line indicates the extent of
the core from the solid vertical line. For computational simplicity,
we do not calculate the region behind the core, as the focus of this
work lies on the penetration depth at the front side.
\label{fig:CLOUDY}
}
\end{figure*}

More realistically, we compute the ionization in the cold core with
CLOUDY, using the same setup as in \S2.1, but this time with an
initial hydrogen density $n_{\rm H}=774\,{\rm cm^{-3}}$ in the
range of $0\sim0.5\,{\rm pc}$, while $n_{\rm H}=7300\,{\rm cm^{-3}}$ from
$0.5\,{\rm pc}$ to faraway, corresponding to the cold medium
placed $0.5\,{\rm pc}$ from the central star, and the warm medium
filling the space in between. Then instead of keeping constant density,
we keep the two media in pressure equilibrium, as the central star
illuminates its surroundings.
In this case, the density of the cold medium is
enhanced rapidly, due to the sudden increase of warm medium temperature
and pressure. After the attenuation of the warm medium, the typical
photon penetration depth inside the cold medium is only about
$0.02 \,{\rm pc}$. This is much smaller than the typical core radius
$r_c \approx 0.1\,{\rm pc}$.
Therefore, only cores very near the massive star or very small cores
are subjected to this effect. The temperature inside the cores further
away is limited by the cooling process to be around 15\textendash20 K, which is
two times higher than before the ignition of the UV flux (see
Figure \ref{fig:CLOUDY(b)},  for ${\rm depth}>0.5\,{\rm pc}$).

In addition, we also investigate the classical evaporation of cores
due to the high temperature environment. Using the formula for the
evaporation rate of clouds embedded in a $10^4$~K gas
\citep{GrahamLanger1973,CowieMcKee1977,McKeeCowie1977}, with the
limitations that the background gas is either fully ionized or neutral,
\begin{equation}
\dot{m} = \frac{16\pi\mu\kappa\,r_c}{25\,k} =
\begin{cases}
1.3\times10^{15}\,T^{1/2}\,r_{\rm c,pc}\,{\rm g}\,{\rm s^{-1}}\, ,
& {\kappa=\kappa_{\rm n}}
\cr 2.75 \times 10^4\,T^{5/2}\,r_{\rm c,pc}\,{\rm g}\,{\rm s^{-1}}\, , & {\kappa=\kappa_{c}}
\end{cases}
\end{equation}
Here, $T$ is the environment temperature, $k$ is the Boltzmann's constant, $r_{\rm c,pc}$ is the radius of
the core in units of pc, $\kappa_{\rm n}$ is the neutral conductivity,
$\kappa_{c}$ is the classical conductivity for a fully ionized gas
\footnote{Note that a different mean molecular weight $\mu$ is used
for the two cases.}. For a typical core with mass $m_{\rm c}$, the evaporation timescale
therefore is $\tau_{\rm evap,n} = m_{\rm c}(\dot{m})^{-1} = 8\times10^5\,r_{\rm c,pc}^2\,{\rm Myr}$ in the neutral
case and $\tau_{\rm evap,i} = 4\times10^8\,r_{\rm c,pc}^2\,{\rm Myr}$ in
the fully ionized case.
Based on these calculations, the classical evaporation only
affects cores with radius as small as $0.01\,{\rm pc}$.

We conclude from the above calculation that the evaporation
effect is negligible. This is understandable as both the recombination
timescale and cooling timescale inside the cold medium are much
shorter than these timescales inside the warm gas due to their density contrast.
Therefore, the ionization fraction and temperature of the cold medium
can be maintained at low levels. We do not include the change of the IMF due
to the evaporation in the calculation below.

\subsection{Change in Bonnor-Ebert mass and star formation timescale}
The star formation timescale in general can be as long as
$100~{\rm Myr}$ \citep{Ostriker2010}. However, the collapse
process is speeded up considerably by the compression of the dense
cores due to the UV feedback \citep{Gritschneder2009MNRAS}. More
importantly, the enhancement of the external pressure reduces
the Bonnor-Ebert mass and induces the formation of low-mass stars.

The Bonnor-Ebert mass  of a cold core can
be expressed as $m_{\rm BE}\propto P_{\rm ext}^{-0.5}T_{\rm int}^{2}$
where $P_{\rm ext}$ is the external pressure (see also equation(4)
in \citealt{Lada2008}). Prior to the formation of a massive star,
the temperature of the warm medium is $T_{\rm ext}\sim100~{\rm K}$,
and the cores' temperature is $T_{\rm int}\sim10~{\rm K}$.  In a
pressure equilibrium, the density contrast between the cores and
medium is $\sim 10$.  The influx of the UV photons from an emerging
massive star ionizes the tenuous warm medium within a pc from the star
and raises the medium's temperature to $T_{\rm ext}\sim9000~{\rm K}$.
In contrast, the temperature within the cores remains at
$T_{\rm int}\sim20~{\rm K}$ (see Fig \ref{fig:CLOUDY}).

The ionization front propagates through the warm medium more
rapidly than the sound speed. Consequently, the increase in
$T_{\rm ext}$ leads to an increase in $P_{\rm ext}$ by a
similar factor before the medium's density can readjust to
a new pressure equilibrium. Due to the combined effect of
the temperature and pressure increase in both
the warm medium and the cold medium, the Bonnor-Ebert
mass decreases by a factor of 2.38, shifting from around
$2~M_\odot$ \citep{Lada2008} to $0.84~M_\odot$ in our model.

Due to this sudden increase of the external pressure, global,
coordinated star formation is induced,  leading to a star burst.
Several factors may introduce some dispersions in the value of
the modified $m_{\rm BE}$.  For example, the flux of ionizing
photons emitted by the first O/B star would be reduced by an
order of magnitude if its mass is halved (the one
we have considered in Figure \ref{fig:CLOUDY} is $40~M_\odot$).  Nevertheless,
the final temperature of the ionized region and the core only
change slightly. The Bonnor-Ebert mass associated with slightly
lower $T_{\rm ext}$ ($8500~{\rm K}$) and
$T_{\rm int}$ ($15~{\rm  K}$) would reduce $m_{\rm BE}$ to
 $0.47~M_{\odot}$.
In this case, the Str{\"o}mgren sphere around the star will be much smaller,
which would  slow down the global star formation process
and extend the age spread in this nebula. If the global
star formation timescale is comparable or longer than the timescale of core coagulation,
the dynamics of cores may affect the IMF further.
Here, we neglect this effect on
the age spread and the effect from the evolution of cores.
Once a core becomes gravitationally unstable, it collapses on a
free fall timescale\footnote{Note that after the collapse
the accretion flow onto the cores could be still active and the
star formation process may continue.  Here, we are mainly
concerned with the epoch before stellar feedback is activated.
The star formation timescale $t_{\rm sf}$ should still be
well approximated with the free fall time $t_{\rm ff}$.}.

\begin{equation}
t_{\rm sf}\sim t_{\rm ff}=\sqrt{\frac{r_{\rm c}^3}
{G\,m_{\rm c}}}\sim 0.3\,{\rm Myr}(\frac{r_{\rm c}}
{0.1 \,\rm pc})^{3/2}(\frac{m_{\rm c}}{2\,M_{\sun}})^{-1/2}.
\label{eq:sf timescale}
\end{equation}
This timescale is much shorter than that associated with
the dynamical evolution of the dense cores prior to the
formation of the first massive stars, which is
about $5~{\rm Myr} $ from H13.  After the external medium
is ionized by the first massive stars, the compression
of the cores reduces the cross section and enhances the density contrast
between the core and the external medium.This increases both the
coagulation and the fragmentation timescales. Therefore, the further
dynamical evolution of the  CMF can be neglected.

\begin{figure*}
\centering
\subfigure[]
{
\epsscale{.40}
\centering
\includegraphics[width=0.45\linewidth]{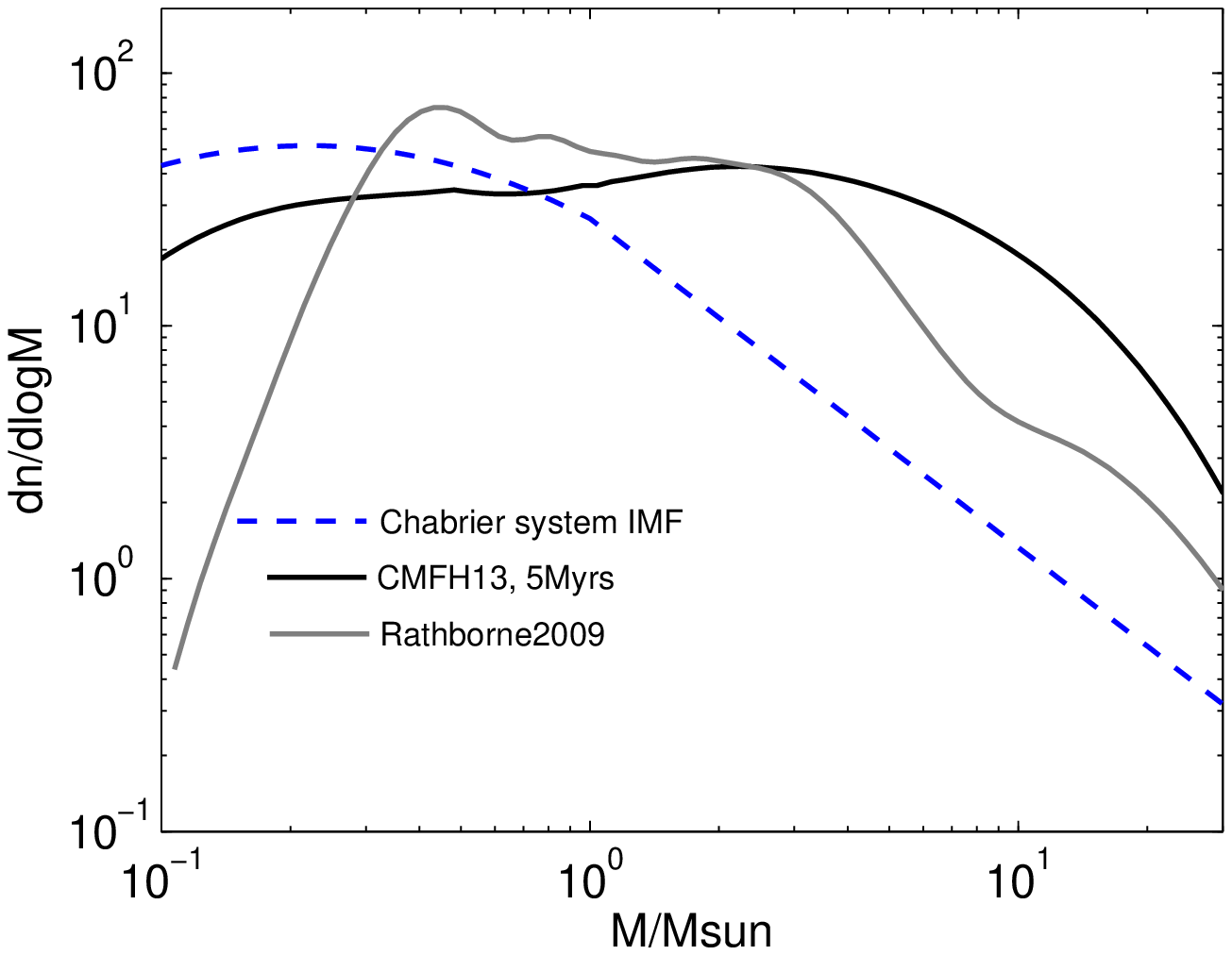}
\label{fig:result(a)}
}
\subfigure[]
{
\epsscale{.40}
\centering
\includegraphics[width=0.45\linewidth]{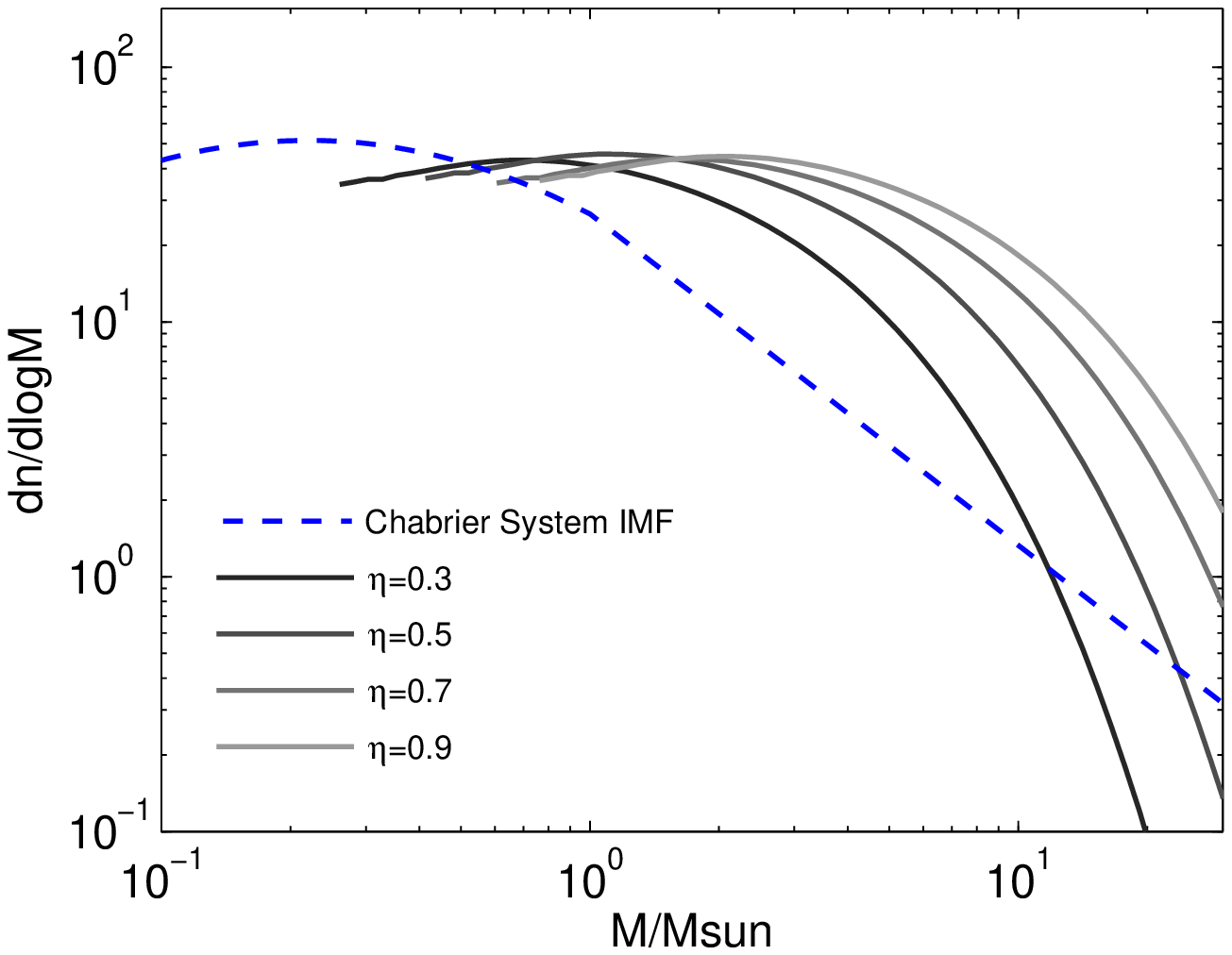}
\label{fig:result(b)}
}\\
\subfigure[]
{
\centering
\includegraphics[width=0.45\linewidth]{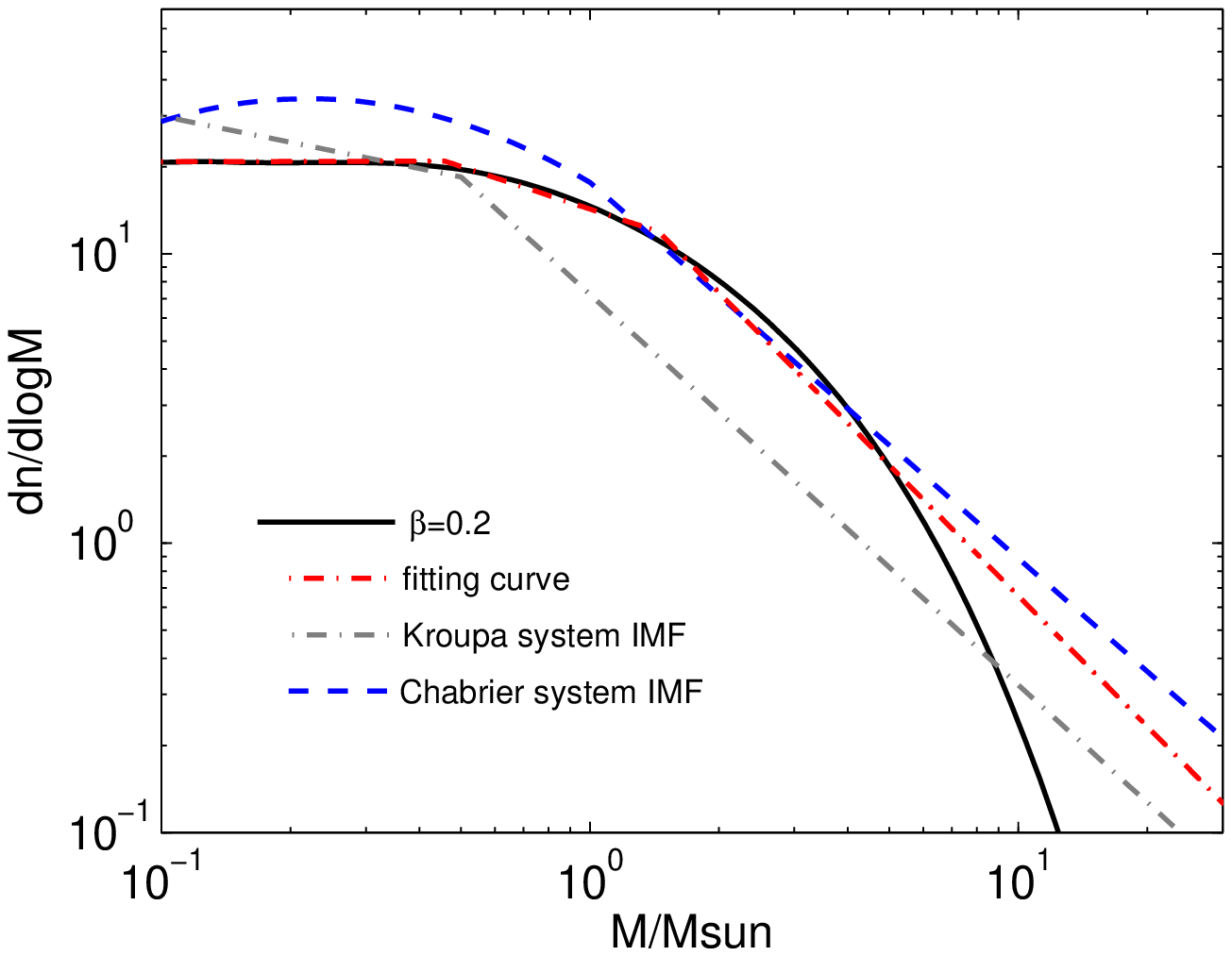}
\label{fig:result(c)}
}
\subfigure[]
{
\centering
\includegraphics[width=0.45\linewidth]{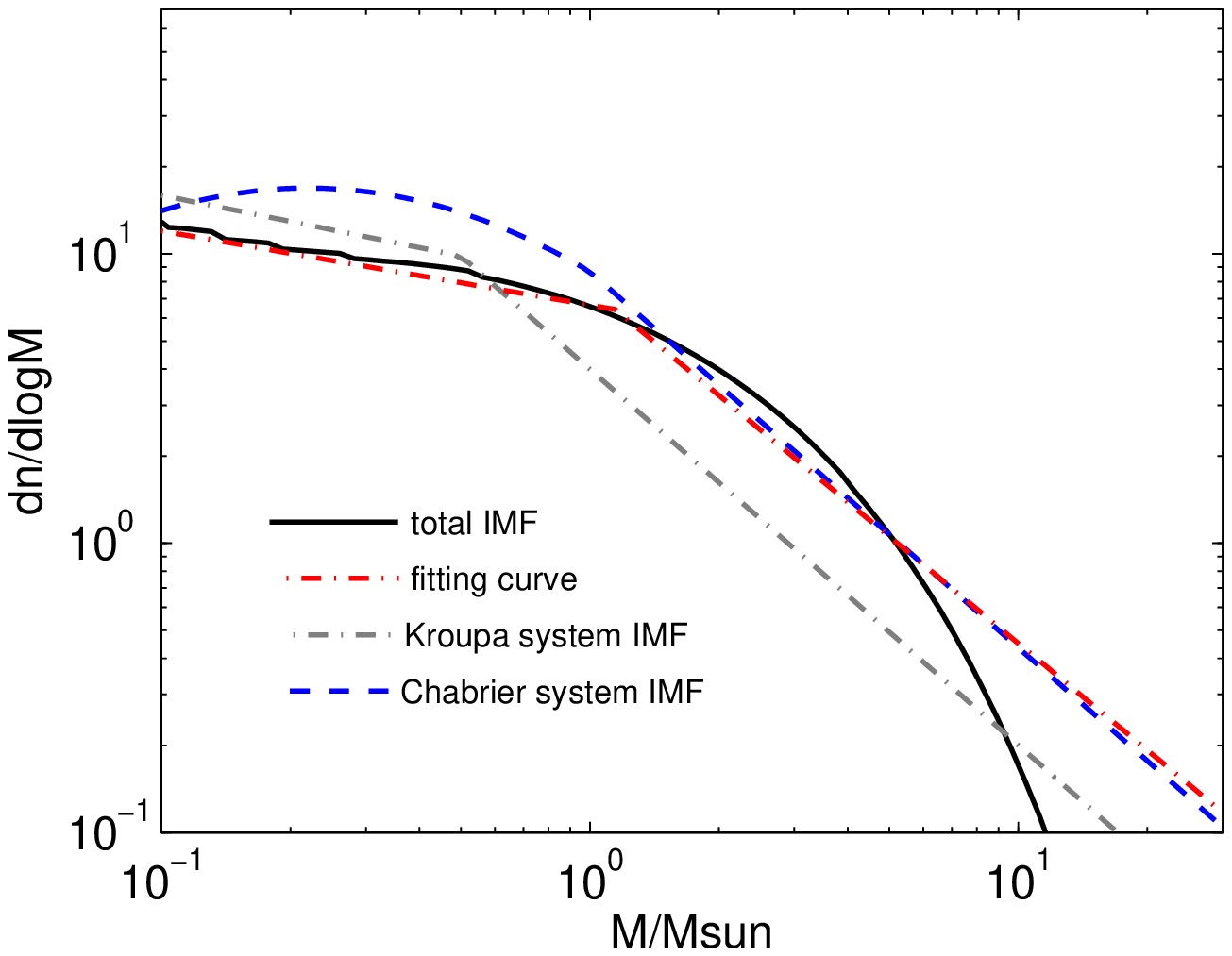}
\label{fig:result(d)}
}

\caption{
CMFs and IMFs generated in our model under different assumptions.
(a) The CMF obtained from H13 (solid black line) is adopted
as the initial condition for our calculations.
Grey line shows the observed CMF in the Pipe from \citet{Lada2009}.
For comparison, the Chabrier IMF \citep{Chabrier03}, shifted to the low mass
end by a factor of 3, is shown as the dashed blue line.
(b) Grey curves are displaying the stellar IMFs for single stars with different
retention factors $\eta_{\rm sf}(m_c)$.
Any core evolution through coagulation or fragmentation is neglected.
(c) The black solid line is showing the generated total IMF if all the cores undergo binary fragmentation
with $\beta = 0.2$ (see Equation \ref{eq:binary})
and $\eta_{\rm sf}(m_c)=0.3$.
(d)The black solid line is showing the total IMF,
assuming $100\%$ binary star formation. The secondary stellar masses
follow a power-law distribution with index  $\alpha = 1.5$
(see Equation \ref{eq:companion}). We show both the Chabrier IMF (blue
dashed curve) and the Kroupa IMF \citep{Kroupa2002}(grey dash-dotted lines) as a comparison. The modeled IMF
is fitted with a broken power-law (red dash-dotted lines).
\label{fig:result}
}
\end{figure*}

\begin{figure}
{
\includegraphics[width=0.9\linewidth]{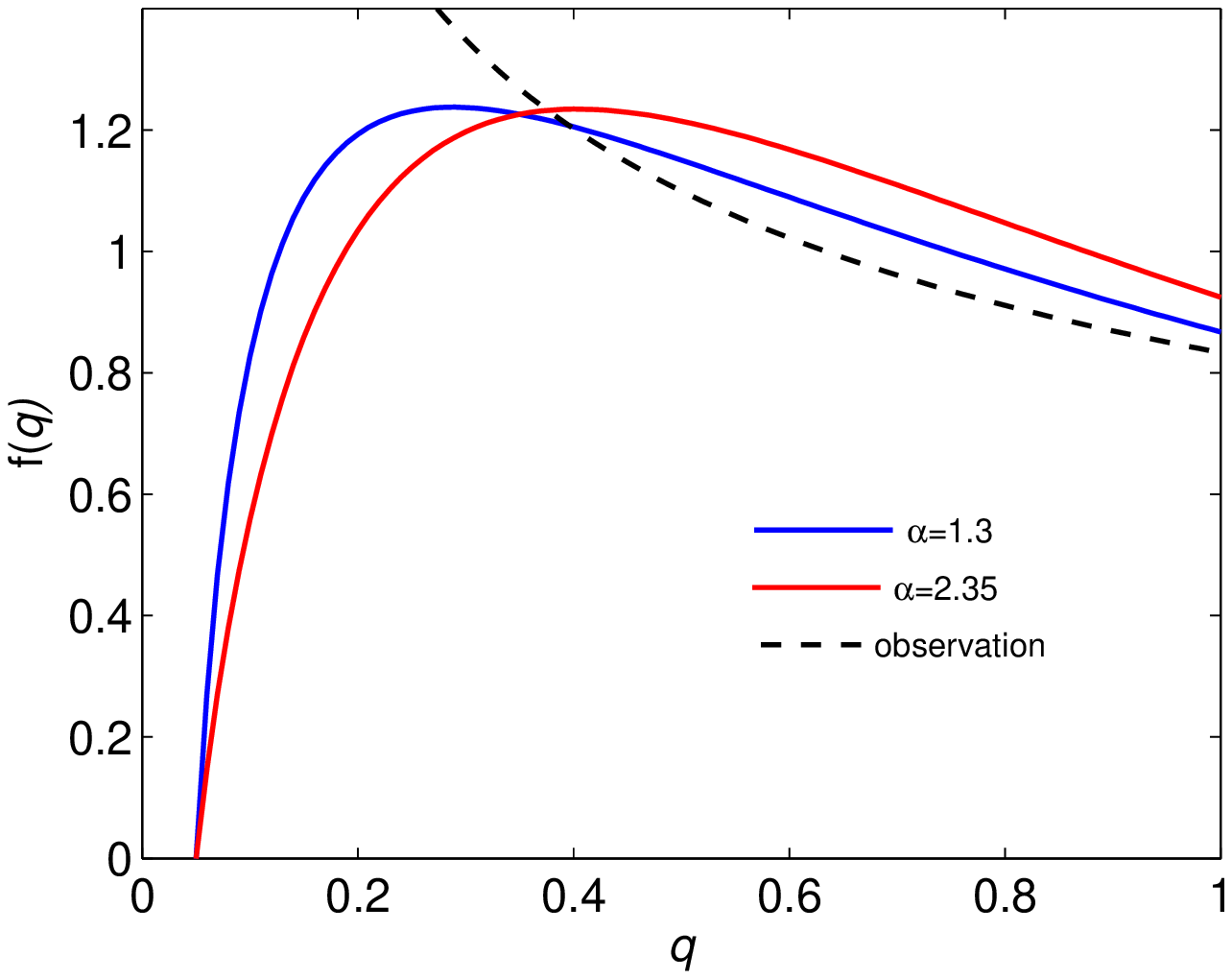}
}
\caption{
\label{fig:analytic}
Analytical results for f(q), i.e., the probability distribution
function (PDF) of observed mass ratios ($\textit{q}=m_{\rm sec}/m_{\rm prim}$) in
binary systems.
The conditional distribution of
mass ratio given the progenitor core mass is assumed to be a uniform distribution
in ($\textit{q}_{\rm min}$,1) regardless of $m_{\rm core}$ (see Eq.\ref{eq:uniform distribution of q}).
For the CMF prescription $dn/dm=m^{-\alpha}$, the
blue line corresponds to $\alpha = 1.3$ and the red line corresponds
to $\alpha = 2.35$. The total accumulated probability is renormalized to
1 and $\textit{q}_{\rm min} = 0.05$. The dashed line is the best fit power-law ($f(\textit{q})\propto \textit{q}^{-0.4}$)
to the observations
from \citet{Thijs2007}.
}
\end{figure}

\section{From the CMF to the IMF}
The initial stellar mass function  is determined by the induced collapse
of the dense cores with a preset CMF. The calculations presented here follow
our previous work on the CMF (H13). In the previous paper, we discussed the evolution
of the dense core mass distribution with a modified coagulation equation.
The CMF acquires a stable shape which resembles a typical observed CMF after
several million years of evolution time. During this stage prior to the triggered
star formation we discuss here, a typical star
formation timescale is around $100~{\rm Myr}$, which keeps the starless
nature of the system. Figure \ref{fig:result(a)} shows the pre-stellar core mass
distribution compared with the Chabrier System IMF
\citep{Chabrier03} and recent observations \citep{Lada2009}. The intermediate mass range of the CMF
($0.3\textendash0.5$~$M_\odot$) can be also parameterized with similar two power-law
slopes as the Kroupa IMF,
only shifted to a higher mass range by a factor of about $3$ \citep{Lada2008}.
We note that both the observed CMF and
modeled CMF have a slightly steeper slope than the IMF at the high mass end.
However,  the uncertainty is also higher in those bins due to small number statistics.
We now take the modeled CMF (H13) as the initial condition and assume the global star
formation is triggered simultaneously. During a time step $\delta t$,
the number of stars in the mass range $m\sim m+dm$ increases by
\begin{equation}
\frac{\Delta{n^{*}(m^{*})}}{\Delta{t}} = \\
\int_{m_{\rm min}}^{m_{\rm max}} m_c n(m_c) \eta_{\rm sf} \gamma_{\rm sf} \\
\frac{P(m_c,m^{*})}{m^{*}} \,dm_c
\label{eq:main}
\end{equation}
where $n(m_c)$ is the number density of cores within the mass interval
$(m_c, m_c+dm_c)$, $\eta_{\rm sf}(m_c)$ is the retention factor (see
section 3.1), $\gamma_{\rm sf}(m_c)$ is the star formation rate
for a dense core with mass $m_c$, and $P(m_c,m^{*})$ is the percentage
of the core mass $(m_c$) transferred into this specific star mass ($m^*$) bin.

The real star formation process may depend on different initial
conditions of the progenitor cores. Here, we present several limiting
cases. To quantify our results, we fit
our computed IMFs with a Chabrier-like function
$\phi(\mu,\sigma,m_0,\gamma)$, where $m_0$ is the transition mass
separating the two parts of the piece-wise Chabrier-like function.
For masses smaller than
$m_0$, the model IMF is described by a lognormal form with mean around
$\mu$ and dispersion $\sigma$; while the distribution at masses larger
than $m_0$ is described by a power-law with index $\gamma$. We require the
distribution to be continuous at $m_0$.
The best fitting parameters are shown in Table \ref{table:chabrier-like fit}.
We compare these fittings with
the Chabrier IMF \citep{Chabrier03} model and discuss the limits of different prescriptions.

\begin{deluxetable*}{cccccc}
\tablecaption{IMF from different star formation models. \label{table:chabrier-like fit}}
\tablecolumns{6}
\tablewidth{0pt}
\tabletypesize{\footnotesize}
\tablenum{1}
\tablehead{\colhead{} & \colhead{$\mu$($M_\odot$)} & \colhead{$\sigma$($M_\odot$)} & \colhead{$m_0$($M_\odot$)} & \colhead{$\gamma$} & \colhead{SSR} \\
\colhead{distribution} & \colhead{mean} & \colhead{dispersion} & \colhead{transition mass} & \colhead{power-law index} & \colhead{Sum of Squared Residuals} }
\startdata
Chabrier   &   0.22   & 0.57  &  1.00  & 1.30  & N/A  \\
single0.3  &   0.66   & 0.54  &  4.05  & 2.19  & 0.000011 \\
single0.5  &   1.05   & 0.54  &  6.42  & 2.18  & 0.000011  \\
single0.7  &   1.55   & 0.54  &  9.36  & 2.14  & 0.000011  \\
single0.9  &   1.95   & 0.55  &  11.02 & 1.98  & 0.0000097  \\
\enddata
\tablecomments{
List of parameters for the resulting IMFs modeled with a
Chabrier type function. We fitted the resulting IMF in the mass range
$0.1\textendash30~M_{\odot}$ with $\frac{dn}{d\log\,m}\propto
e^{\frac{-{\log(m/\mu)}^2}{2\sigma^2}}$ for $m<m_0$, and with
$\frac{dn}{d\log\,m}\propto m^{-\gamma}$ for $m>m_0$.
Continuity is imposed at $m_0$.
Columns starting with `single' refer to the results from the single star formation
model in section 3.1, and the number attached to it corresponds
to the retention efficiency.
}
\end{deluxetable*}

\subsection{Case 1: burst of single stars}
Given the similarity between the CMF and the IMF, some authors suggested an
one-to-one conversion of dense cores into young stars \citep[e.g.,][]{Lada2008}.
Although this scenario has been criticized on the basis of binary
stars' prevalence \citep[see e.g.,][]{Smith2009}, it is nonetheless
informative to explore this simplest possibility. In this case,
$P(m_c, m^*) = \delta(m^* - m_c \eta_{\rm sf}(m_c))$ (see Equation
\ref{eq:main}). $\eta_{\rm sf}$ is the retention factor, which
is the fraction of core mass finally remained in the stars, for one particular core.
While the more often loosely used phrase `star formation efficiency'
should refer to the global ratio
of total stellar masses over total progenitor core masses, in a certain region and time-scale.
In Figure \ref{fig:result(b)} we show the IMFs generated
with this model. We assume only cores with mass exceeding the
Bonnor-Ebert mass can form stars, such that the star formation rate

\begin{equation}
\gamma_{\rm sf}(m_c)=
\begin{cases}
0\ & {\rm if}\ m_c<m_{\rm BE}
\cr\Gamma_{\rm sf} & {\rm if}\ m_c \ge m_{\rm BE}\cr
\end{cases}
\label{eq:sf rate}
\end{equation}
Here, $\Gamma_{\rm sf} = t_{\rm sf}^{-1}$ (see Equation \ref{eq:sf timescale})
is the characteristic star formation
rate. We also assume that the retention factor, i.e. the amount of the core mass
retained in the resulting star, is assumed to be constant with a value
$0<\eta_{\rm sf}(m_c)<1$.

The results in Figure \ref{fig:result(b)}
show that the IMFs are shifted toward the low-mass end, with a broad peak near
the original Bonnor-Ebert mass modified by the retention factor.
With $\eta_{\rm sf}=0.3$ the new peak is at around $0.7 ~M_\odot$,
which corresponds to Lada's suggestion \citep{Lada2008}.
The overall shape of the IMF resembles the observed stellar IMF. The
low-mass end have similar dispersion with that of
Chabrier's IMF, provided the retention factor $\eta_{\rm sf}$ is
not a sensitive function of the progenitor core mass. The high-mass end
power-law is steeper than the Salpeter slope \citep{Salpeter1955},
where the observational uncertainty in this range is also relatively large.
However, the minimum mass of stars we can produce with a single star formation
prescript, is dependent on the critical Bonnor-Ebert mass,
and $\eta_{\rm sf}$ such that
$m_{\rm cut-off}\sim m_{\rm BE,new}\eta_{\rm sf}$.
Due to this cut-off even with a low value of $\eta_{\rm sf}$ ($\eta_{\rm sf}=0.3$, dark grey
line in Figure \ref{fig:result(b)}), this model has no inference for the lowest mass stars.
 The bulk of the modeled IMFs are more
massive than the Chabrier IMF. We conclude that a non-unity retention
factor alone would not explain the transition from CMF to IMF.

\subsection{Case 2: Solely binary formation}
A large fraction of the young stars are binaries. Each component
of the binary contributes to the statistics of the IMF.
Another limiting case of interest is the possibility
that all the cores with mass in excess of the modified Bonnor-Ebert
mass would fragment into binary stars. In principle, binary star
formation is a consequence of rotational fragmentation and the
kinematic properties are determined by the angular momentum distribution
of the cores and their cooling ability. However, we are primarily
interested in the IMF rather than the period distribution. We adopt
an idealized power-law distribution for the mass ratio $\textit{q}$ such that
\begin{equation}
\frac{dn^*_{\rm binary}}{d\textit{q}}\propto \textit{q}^{-\beta}.
\label{eq:binary}
\end{equation}
Here, the cut-off is taken as $\textit{q}_{\rm min}=0.05$.
We explore a range of power-law indices $\beta$ between $0.2$ and $2.5$,
which essentially covers the extreme limits.
We assume a constant retention $\eta_{\rm sf}=0.3$
and star formation rate same as in Equation (\ref{eq:sf rate}). The
results here remain valid if the retention factor
does not vary strongly with the progenitor core mass.

All the resulting IMFs with different $\beta$ display a broad shifted
peak around the new, reduced Bonnor-Ebert mass (Figure \ref{fig:result(c)}).
The results are very slightly affected by different $\beta$ in this wide range,
so only $\beta=0.2$ is shown.
The mass associated with the turn-over of the
IMF is smaller than that found for the single star formation
model. The segments in $0.46~M_\odot<m\le 1.45~M_\odot$ with power-law index $\gamma_1=0.489$ and in $1.45~M_\odot<m\le 10~M_\odot$ with power-law index $\gamma_2=1.5$. The two indices are
 quite similar to those of the piece-wise Kroupa IMF.
We expect that the allowance of trinaries and quaternaries will further move
the peak towards low-mass end.

\subsection{Case 3: The binary companion's IMF}

In a classical study of the binary star population census by \citet{Mayor1991},
the IMF for the companions of G-dwarf stars is thoroughly analyzed.
They show that binaries on average can be formed by random combination
of stars drawn from the same IMF.
Following their basic approach, we consider the IMF of
the primary and secondary stars separately.
We assume all the progenitor cores fragment into binaries and the secondary
star masses follow a distribution (e.g. Gaussian, Salpeter or Miller-Scalo \citep{Miller-Scalo} power-law)
independent of their
primaries' mass. For simplicity, we adopt a power-law, so the transfer
functions as in Equation (\ref{eq:main}) become:

\begin{equation}
P_{\rm sec}(m_{\rm sec},m_c)\propto dn_{\rm sec}/dm_{\rm sec}\propto (m_{\rm sec})^{-\alpha},
\label{eq:companion}
\end{equation}

\begin{equation}
P_{\rm prim}(m_{\rm sec},m_c) = P_{\rm sec}(\eta m_c-m_{\rm sec},m_c),
\end{equation}
where the subscript 'sec' and 'prim' represent the secondary
and the primary star. For $P_{\rm sec}$ the range of
$m^*$is $0.08~M_\odot < m_{\rm sec} < 0.5~\eta_{\rm sf} m_c$.
The power-law index for the secondary
star mass distribution, $\alpha$, varies from $0.8$ to
$1.5$ in our calculations, while
the star formation rate is as in Equation (\ref{eq:sf rate}) and the
retention factor $\eta_{\rm sf} = 0.3$ is still assumed.
Here, the IMF of the primary star preserves its
characteristic broad peak and overall shape. The characteristic mass
associated with this peak is smaller than that of the CMF.

The consequent IMF for the primary
stars is very slightly influenced by $\alpha$ in this wide range, so we only show
the result with $\alpha = 1.5$ in Figure \ref{fig:result(d)}. Following
\citet{Mayor1991}, we compare the total (primary and secondary) simulated
IMF with some well-known IMFs, such as the Kroupa IMF.  The shape of the modeled IMF resembles
the Kroupa IMF more closely at masses below $0.7~M_{\odot}$. We fit it with two broken
power-laws. The low-mass end ({\bf{$m<1 ~M_\odot$}})
completely reproduces
the assumed power-law we put in, with $dn/d\log\,m = m^{-0.5}$, while the
intermediate mass range ($1 ~M_\odot <m<10 ~M_\odot$) recover the
Salpeter power-law $dn/d\log\,m = m^{-1.3}$.

\subsection{Analytical results for the binary mass ratio distribution}

An alternative approach to characterize binary star statistics is to
utilize the mass ratio, defined as
$\textit{q}=m_{\rm sec}/m_{\rm prim}$. Observations show that the mass ratio also roughly
follows a power-law distribution, as in Scorpius OB2 for intermediate
mass stars \citep{Thijs2007}.
Given a probability distribution of mass ratio
$p(\textit{q}|m_{\rm core})$ in a binary formation process (formed
from cores with mass $m_{\rm core}$), we can combine it with
the CMF $p(m_{\rm core})$ to predict the mass ratio distribution
$f(\textit{q})$:
\begin{equation}
p(\textit{q})=\int_{m_{\rm min}}^{m_{\rm max}}p(\textit{q}|m_{\rm core})
p(m_{\rm core})dm_{\rm core}.
\end{equation}
With a power-law mass distribution of the progenitor dense cores and
an assumed constant retention, we can estimate the
probability of finding a binary system with stellar masses of $m_{\rm prim}$
and $m_{\rm sec}$ to be
\begin{equation}
f(m_{\rm prim},m_{\rm sec})\propto (m_{\rm prim} + m_{\rm sec})^{-\alpha},
\end{equation}
where $\alpha$ is the power-law index of the progenitor CMF. For random
pairing of binary systems \citep{Mayor1991},
\begin{equation}
p\text{\textit{q}}|m_{\rm core})=\frac{1}{1-\text{\textit{q}}_{\rm min}} \ \ \ \
\:{\rm for} \text{\textit{q}}\in [\text{\textit{q}}_{\rm min},1]
\label{eq:uniform distribution of q}
\end{equation}
with $\textsl{\textit {q}}_{\rm min}$ as the minimum allowed binary ratio.

The cumulative probability $P(\textit{q}<x)$ for
all the binary systems would then be
\begin{equation}
P(\textit{q}<x) = \int_{m_{\rm min}}^{x m_{\rm max}}\,dm_{\rm prim} \\
\int_{\frac{m_{\rm prim}}{x}}^{m_{\rm max}}\,dm_{\rm sec}f(m_{\rm prim},m_{\rm sec}).
\end{equation}
Thus, the statistically averaged distribution of mass ratios $\textit{q}$ would be
\begin{equation}
f(\textit{q}) \equiv \frac{dP(\textit{q})}{d\textit{q}}\propto \frac{1+(2-\alpha)\textit{q} + (1-\alpha)\textit{q}_{\rm min}^{2-\alpha} \textit{q}^{\alpha-2}}{(2-\alpha)(1+\textit{q})^\alpha}\\
- (1+\textit{q})^{1-\alpha},
\end{equation}
where $m_{\rm max} (=\eta_{\rm sf} m_{core})$
is the upper limit for progenitor core mass.
We choose $\alpha$ values corresponding to Kroupa's IMF power-law indices.
Although there are some uncertainties in the minimum allowed mass ratio,
binary systems with q ratio around 0.05 or less has been observed
so we adopt $\textit{q}_{\rm min} = 0.05$ in the calculation.
Results are shown in Figure \ref{fig:analytic}
and compared with the observed power-law from \citet{Thijs2007}.
Although the distribution of q depends on the value of $\textit{q}_{\rm min}$, the power-law
slope converges, for $\textit{q}>>0.05$,
to the observed value.

\section{Discussion and Conclusions}

In this work, we continue our investigation on the origin of the stellar
IMF. Based on the similarity between the observed stellar IMF and the CMF
of dense starless cores in the Pipe nebula, we assume that they are closely
connected.  In \citet{GritschnederLin2012}, we suggest that the cold cores
of molecular gas are the byproducts of a thermal instability or
the fragmentation of the a shocked shell and they are pressure confined
by tenuous warm atomic medium.  The CMF of these cores is the natural
outcome of their collisional coagulation and their fragmentation due to
their hydrodynamic interaction with the surrounding medium (H13).
We also assume that these cores become unstable, undergo gravitational
collapse, and evolve into protostars after their mass exceeds the
Bonnor-Ebert mass.

Although this simple model reproduces the basic observed slopes of
the stellar IMF, there is a factor-of-three offset between the stellar
mass associated with the peak of the stellar IMF and that associated
with the peak of the cores' CMF.  The main motivation of the investigation
presented in this paper is to suggest a mechanism to account for this
shift.

In \citet{Huang2013}, we suggest that the peak of the CMF is essentially
the Bonnor-Ebert mass of the cores.  In typical molecular clouds such as the
Pipe nebula, the Bonnor-Ebert mass is $> 1~M_\odot$.
In principle, cores less massive than the Bonnor-Ebert mass are stable
and they do not turn into stars.  Yet, many low-mass stars are formed
in young stellar clusters.

In this work, we make an attempt to resolve the issues of 1) the offset
between the mass associated with the peak of the CMF and the IMF, and 2) the
prolific production of sub-solar type stars in molecular clouds.  We
demonstrate that the onset of first massive stars in these clouds
photoionize and heat the surrounding medium without significantly
changing the ionization fraction and temperature of the cores.  This
feedback effect largely increases the external pressure which confines
the cores (by up to two orders of magnitude).  This change leads to
the compression of the cores and a reduction in their Bonnor-Ebert mass.

The collapse of cores with a mass greater than the modified
Bonnor-Ebert mass (and less than the original Bonnor-Ebert mass) can
now lead to the formation of a large population of sub-solar-mass stars.
The stellar IMF generally preserves the shape of the Salpeter-like CMF
with a significant lowering in the peak mass. The shape of
the IMF (dispersion and high mass end slope) is not strongly modified
by the formation of binary stars through rotational fragmentation.
In our model, the inclusion of binary fragmentation during the collapse
is essential for the production
of stars with mass lower than the modified Bonnor-Ebert mass.

One implication of this induced star formation scenario is that the
intrinsic age spread of the stellar cluster is naturally very small.
In our analysis, we have adopted an idealized treatment of the
retention efficiency (often loosely referred to as star formation
efficiency). A mass dependence in the retention factor may lead to
some modification in the IMF.  We present several models
for single and binary star populations. They generally
reproduce the transition from the CMF to the IMF suggested by
the observational data. A prolific production of triple and hierarchical
systems may also lead to the formation of very low-mass stars.

Finally, these models are constructed with solar composition.  In
metal-deficient gas, such as protoglobular cluster clouds, the
inability to cool may lead to a higher internal temperature and a higher
Bonnor-Ebert mass in the cores. This feedback mechanism may be
particularly important in triggering the formation of low-mass stars
with lifespans in excess of 10 Gyr.  We will further explore this
possibility in the future.

\section{Acknowledgements}
We thank M.B.N. Kouwenhoven and C. Lada for useful conversations.
DNCL acknowledges support by NASA through NNX08AL41G. MG acknowledges
support from the Humboldt Foundation in form of a Feodor Lynen Fellowship.

\clearpage
\end{document}